\documentclass[usenatbib,onecolumn]{mn2e}
\usepackage{epsfig}

\newcommand{\vect}[1]{\mathbf{#1}}
\newcommand{\textok}[1]{#1}

\title[Reconstructing Galactic Collisions]{Identikit 2:
  An Algorithm for Reconstructing Galactic Collisions}

\author[J. E. Barnes]{Joshua E. Barnes\\
  Institute of Astronomy, University of Hawaii, 
  2680 Woodlawn Drive, Honolulu, HI 96822, USA}

\begin{document}

\maketitle

\begin{abstract}
Using a combination of self-consistent and test-particle techniques,
Identikit~1 provided a way to vary the initial geometry of a galactic
collision and instantly visualize the outcome.  Identikit~2 uses the
same techniques to define a mapping from the current morphology and
kinematics of a tidal encounter back to the initial conditions.  By
requiring that various regions along a tidal feature all originate
from a single disc with a unique orientation, this mapping can be used
to derive the initial collision geometry.  In addition, Identikit~2
offers a robust way to measure how well a particular model reproduces
the morphology and kinematics of a pair of interacting galaxies.  A
set of eight self-consistent simulations is used to demonstrate the
algorithm's ability to search a ten-dimensional parameter space and
find near-optimal matches; all eight systems are successfully
reconstructed.
\end{abstract}


\section{Introduction}

Dynamical modeling of specific pairs of interacting galaxies is a
subject with considerable history.  \citet[][hereafter TT72]{TT72}
bolstered their interpretation of bridges and tails as tidal features
by presenting test-particle models of Arp~295, M~51, NGC~4676, and
NGC~4038/9; the power of such models was demonstrated when
\citet{S74} confirmed TT72's prediction for the relative velocities of
the two galaxies making up NGC~4676.  As observational and numerical
techniques have improved, modeling of interacting systems has
generated an extensive literature (see \citealt{BH09}, hereafter BH09,
for a partial list).  The motivation for dynamical modeling has
evolved over time.  While early studies focused on testing the tidal
theory of galactic encounters, more recent work has used dynamical
modeling to help interpret observations, probe the structure and
dynamics of unseen matter, and reconstruct the dynamical histories of
merging galaxies.

Despite the advances of the past few decades, it remains \textok{a
challenge} to create models matching the detailed morphology and
kinematics of a pair of colliding galaxies.  One fundamental
\textok{complication} is the inherent uncertainty in inferring the
distribution \textit{and} dynamics of dark matter strictly by its
effects on luminous material.  Another is the \textok{violent
reprocessing} of interstellar material -- including rapid star
formation -- in galaxy collisions.  However, there are three rather
technical issues which also limit progress:
\begin{enumerate}
\renewcommand{\theenumi}{(\arabic{enumi})}
\item
A galactic collision is described by a large number of parameters
which interact in highly non-linear ways.
\item
Simulating galactic collisions is computationally intensive.
\item
The criteria for a successful match are not easily translated into
quantitative terms.
\end{enumerate}

\begin{figure}
\begin{center}
\includegraphics[clip=true,width=0.25\columnwidth]{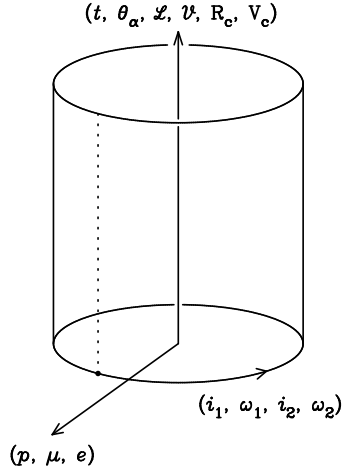}
\caption{An abstract representation of the sixteen-dimensional
parameter space of galaxy interactions.  The radial coordinate
represents the initial orbit, the azimuthal coordinate represents the
disc orientations, and the vertical coordinate represents the
parameters chosen after a simulation is run. A conventional N-body
simulation explores the parameter subspace represented by the dotted
line, while a single Identikit simulation explores the entire
cylindrical surface.  \label{parameters}}
\end{center}
\end{figure}

The parameters necessary to simulate an encounter of two disc galaxies
fall into three groups, as illustrated in Fig.~\ref{parameters}.  The
first group specifies the initial orbits of the galaxies; assuming
these orbits are asymptotically Keplerian at early times, the required
parameters are the periapsis separation $p$, the orbital eccentricity
$e$, and the mass ratio $\mu$.  The second group describes the spin
vector of disc $d$ (where $d = {1}, {2}$) with respect to the angular
momentum of the relative orbit and the separation vector between the
galaxies at periapsis; this vector is parametrized by the inclination
$i_{d}$ and argument to periapsis $\omega_{d}$ of each disc.  Together
with any parameters needed to describe the internal structures of the
two galaxies, the first and second groups specify the \textit{initial
conditions} for a galactic encounter.  The third group consists of the
time $t$ since first periapsis, and parameters which map the
simulation onto the observational plane: three Euler angles
$\theta_\alpha$ specifying the viewing direction, scaling factors
$\mathcal{L}$ and $\mathcal{V}$ \textok{which transform dimensionless
simulation positions and velocities, respectively, into real physical
quantities}, and the centre-of-mass position on the plane of the sky
$\vect{R}_\mathrm{c}$ and radial velocity $V_\mathrm{c}$.  These
parameters may be chosen \textit{after} a simulation has been run.

Of these sixteen parameters, only a few have a~priori constraints.
TT72 argued that the orbital eccentricity should be $e \simeq 1$; this
is generally supported by cosmological simulations \citep{KB06},
although the $e$ distribution extracted from these simulations
includes a tail to $e < 1$.  The mass ratio $\mu$ may be estimated
from the relative luminosities of the two galaxies -- provided that
the galaxies are still distinct and that interaction-induced star
formation has not significantly altered their luminosities.  Finally,
the scale factors $\mathcal{L}$ and $\mathcal{V}$ are are not
completely arbitrary since the pre-encounter galaxies should have
radii and circular velocities comparable to those of other disc
galaxies.

\subsection{Identikit 1}

Identikit simulations combine test-particle and self-consistent
techniques (BH09).  Each galaxy is modeled by an initially spherical
configuration of massive particles with cumulative mass profile
$m(r)$, in which is embedded a spherical swarm of massless test
particles on initially circular orbits \textok{with angular momenta
uniformly distributed over all directions}.  Two such models are
launched towards each other with orbital parameters $(p, \mu, e)$.
During the ensuing encounter, the massive components interact
self-consistently, approximating the time-dependent potential and
orbit decay of a fully self-consistent galactic collision.  The test
particles mimic the tidal response of embedded discs with all possible
spin vectors; once such a simulation has been run, selecting the
appropriate subset of test particles yields a good approximation to
the tidal response of any particular disc.

In the simplest Identikit implementation, the test particles initially
populating each galaxy model have the same radial distribution as the
discs they are intended to mimic.  Each test particle $i$ is
associated with a normalized vector $\vect{s}_i \in \mathsf{S}^2$
which records the direction of the particle's \textit{initial} angular
momentum with respect to the centre of its parent galaxy.  An
Identikit simulation yields a Monte Carlo representation of an
`extended' distribution function,
\begin{equation}
  g_{d}(\vect{r}, \vect{v}, \vect{s}; t) \doteq
  \sum\nolimits_{\textstyle i}
    \delta^3(\vect{r} - \vect{r}_i(t)) \,
    \delta^3(\vect{v} - \vect{v}_i(t)) \,
    \delta^2(\vect{s} - \vect{s}_i) \, .
\end{equation}
This function gives the phase-space number density at time $t$ of test
particles from disc ${d}$ with position $\vect{r}$, velocity
$\vect{v}$ which initially had angular momentum direction $\vect{s}$.
Here and throughout, `$\doteq$' is used throughout to indicate an
explicit Monte Carlo expression.  On the right-hand side,
$\vect{r}_i(t)$ and $\vect{v}_i(t)$ are the position and velocity of
test particle $i$ at time $t$; these are understood to \textok{also}
depend on the orbital parameters $(p, \mu, e)$.  Once this extended
function has been constructed, the distribution function $f_{d}$ for a
disc with a specific initial spin $\vect{s}_0$ is estimated by
\begin{equation}
  f_{d} (\vect{r}, \vect{v}; t) \propto
  \int\nolimits_{\textstyle \tilde{\vect{s}}_0} d^2\vect{s} \,
    g_{d}(\vect{r}, \vect{v}, \vect{s}; t) \doteq
  \sum\nolimits_{\textstyle \vect{s}_i \in \tilde{\vect{s}}_0} \,
    \delta^3(\vect{r} - \vect{r}_i(t)) \,
    \delta^3(\vect{v} - \vect{v}_i(t)) \, ,
\end{equation}
where $\tilde{\vect{s}}_0 = \{ \vect{s} \in \mathsf{S}^2 \,|\,
\vect{s} \cdot \vect{s}_0 \ge 1-\sigma \}$ and $\sigma \ll 1$ is a
tolerance parameter which determines the solid angle contributing to
the estimate of $f_{d}$.

The reason why $f_{d}$ is estimated by integrating over a finite solid
angle $\tilde{\vect{s}}_0$ is not obvious; it may seem enough to
simply evaluate $g_{d}(\vect{r}, \vect{v}, \vect{s}_0; t)$.  However,
$g_{d}$ and $f_{d}$ are both represented in a Monte Carlo fashion.  To
sample $f_{d}$ well enough for a visual comparison with observational
data requires a few thousand particles; if $g_{d}$ is represented by
$N_\mathrm{test} \sim 10^5$ to~$10^6$ test particles per galaxy, this
requires $\sigma \simeq 10^{-2}$.

This simple version of Identikit has some drawbacks.  Only a small
percentage of the test particles are initially placed at large radii
where they are responsive to tidal forces; \textok{this wastes
computer time}.  \textok{Moreover}, simulated discs \textok{defined by
$\vect{s}_i \cdot \vect{s}_0 \ge 1 - \sigma$} have scale heights which
increase linearly with radius \textok{and look unrealistic when viewed
edge-on}.  Both of these flaws can be addressed by radially biasing
the distribution of test particles.  Following BH09, the test particle
density is multiplied by a factor of $r^2$, and each test particle $i$
is given a weight $\xi_i =
\mathrm{max}(r^\mathrm{init}_i/r_\mathrm{min}, 1)^{-2}$, where
$r^\mathrm{init}_i$ is the initial orbital radius of particle $i$, and
$r_\mathrm{min}$ is a small cut-off radius.  The extended distribution
function is then
\begin{equation}
  g_{d}(\vect{r}, \vect{v}, \vect{s}, \xi; t) \doteq
  \sum\nolimits_{\textstyle i} \delta^3(\vect{r} - \vect{r}_i(t)) \,
                  \delta^3(\vect{v} - \vect{v}_i(t)) \,
                  \delta^2(\vect{s} - \vect{s}_i)
                  \delta(\xi - \xi_i) \, ,
\end{equation}
and the expression for $f_{d}$ becomes
\begin{equation}
  f_{d} (\vect{r}, \vect{v}; t) \propto 
  \int d\xi
    \int\nolimits_{\textstyle \tilde{\vect{s}}_0(\xi)} d^2\vect{s} \,
      g_{d}(\vect{r}, \vect{v}, \vect{s}, \xi; t) \doteq
  \sum\nolimits_{\textstyle \vect{s}_i \in \tilde{\vect{s}}_0(\xi_i)} \,
    \delta^3(\vect{r} - \vect{r}_i(t)) \,
    \delta^3(\vect{v} - \vect{v}_i(t)) \, ,
\end{equation}
where $\tilde{\vect{s}}_0(\xi) = \{ \vect{s} \in \mathsf{S}^2 \,|\,
\vect{s} \cdot \vect{s}_0 \ge 1 - \sigma\xi \}$.

\section{Deriving Disc Spins}
\label{deriving_spins}

Identikit~1 uses the extended distribution function $g_{d}$ in a
`forward' mode which allows a rapid exploration of the outcome for any
choice of initial disc spin.  However, the same function can also be
used in an `inverse' mode, in which the fact that a tidal feature
populates certain regions of phase space can be used to solve for the
initial spin vector of the disc it came from.  Let $\vect{q} =
(\vect{r}, \vect{v}) \in \mathsf{R}^6$ be a point in phase space, and
$\tilde{\vect{q}} \subset \mathsf{R}^6$ be a region of phase space
which is populated by tidal material from disc ${d}$.  Define a
density on the sphere of spin directions $\mathsf{S}^2$:
\begin{equation}
  \Omega_{d}(\vect{s}; \tilde{\vect{q}}, t) =
  \int d\xi
    \int\nolimits_{\textstyle \vect{q} \in \tilde{\vect{q}}}
      d^3\vect{r} \, d^3\vect{v} \,
        \xi g_{d}(\vect{r}, \vect{v}, \vect{s}, \xi; t) \doteq
  \sum\nolimits_{\textstyle \vect{q}_i(t) \in \tilde{\vect{q}}} \,
    \xi_i \delta^2(\vect{s} - \vect{s}_i) \, ,
  \label{omega_def}
\end{equation}
where $\vect{q}_i(t) = (\vect{r}_i(t), \vect{v}_i(t))$.  This function
describes the inverse image of the material populating the phase space
region $\tilde{\vect{q}}$ at time $t$ mapped back onto the sphere
of initial spin directions.  As will be seen shortly, the inverse
image of a region $\tilde{\vect{q}}$ may span a wide range of
initial disc spins $\vect{s}$, and the function $\Omega_{d}$ can have
a rather complicated behavior.  Nonetheless, $\Omega_{d}$ usually has
compact support, meaning that only a subset of all possible disc spins
can produce tidal features populating $\tilde{\vect{q}}$.

Now consider a set of $n_\mathrm{reg}$ phase-space regions
$\tilde{\vect{q}}_j$ (where $j = 1, \dots, n_\mathrm{reg}$) which
sample a given tidal structure.  For example, imagine that disc ${d}$
has produced a tidal tail, and that the regions are distributed along
this tail.  Each of these regions $\tilde{\vect{q}}_j$ can be
populated by some range of initial spins, defined by the set of
$\vect{s}$ for which $\Omega_{d}(\vect{s}; \tilde{\vect{q}}_j, t)
> 0$.  However, the tail as a whole is presumably populated by one
disc with a \textit{unique} spin $\vect{s}_0$, and this spin
\textit{must} lie within the intersection of the images of
the individual regions $\tilde{\vect{q}}_j$.

To find this intersection numerically, it's convenient to work with a
smoothed and normalized version of $\Omega_{d}$.  Define
\begin{equation}
  \overline{\Omega}_{d}(\vect{s}; \tilde{\vect{q}}, t) =
  K(\tilde{\vect{q}}) \int d^2\vect{s}' \,
    w(\vect{s} - \vect{s}') \, 
    \Omega_{d}(\vect{s}'; \tilde{\vect{q}}, t) \doteq
  K(\tilde{\vect{q}})
    \sum\nolimits_{\textstyle \vect{q}_i(t) \in \tilde{\vect{q}}} \,
    \xi_i w(\vect{s} - \vect{s}_i) \, ,
  \label{omegabar_def}
\end{equation}
where $w(\Delta\vect{s})$ is a smoothing function with compact
support, and the factor $K(\tilde{\vect{q}})$ normalizes the maximum
value of $\overline{\Omega}_{d}(\vect{s}; \tilde{\vect{q}})$ to unity.
Smoothing replaces the pointillistic Monte Carlo representation of
$\Omega_{d}$ with a continuous version in which the value of
$\overline{\Omega}_{d}(\vect{s})$ reflects the density of points near
$\vect{s}$; accordingly, the radius of the smoothing function should
be somewhat larger than the mean separation on $\mathsf{S}^2$ between
the points.  The advantages of normalization will be illustrated
later.

Given smoothed functions for a set of regions $\tilde{\vect{q}}_j$
tracing the tidal features of a single disc, define the product
function
\begin{equation}
  \Omega^*_{d}(\vect{s}; t) =
    \prod\nolimits_{\textstyle j = 1}^{\textstyle n_\mathrm{reg}}
      \overline{\Omega}_{d}(\vect{s}; \tilde{\vect{q}}_j, t) \, .
  \label{omegastar_def}
\end{equation}
This function may vanish everywhere on $\mathsf{S}^2$, it may have a
complicated structure, perhaps with multiple peaks, or it may have a
simple structure with just one peak.  If it vanishes everywhere, then
\textit{no} single disc spin can populate \textit{all} of the target
regions $\tilde{\vect{q}}_i$; this may indicate that model parameters
such as the time $t$ since periapsis are incorrect.  If it has a
complicated structure, then there are many initial configurations
which can populate all of the target regions.  But if $\Omega^*_{d}$
has a single peak at $\vect{s}_0$ and vanishes everywhere else except
for a small surrounding neighborhood, then the true spin of the disc
must be close to $\vect{s}_0$ -- provided that $t$ and other model
parameters are accurately known.

\subsection{Observational constraints}

Observations of distant galaxies can constrain only some components of
phase-space.  Let the observations of a pair of interacting galaxies
be described using coordinates $\vect{R} = (X,Y,Z)$, with the $Z$ axis
coinciding with our line of sight to the system, so that $(X,Y)$ is
parallel to the plane of the sky.  For simplicity, assume that the
origin of these coordinates lies near the system's centre of mass.
Given a distance to the system, the $(X,Y)$ coordinates may be
expressed in physical units.  Likewise, describe velocities using
coordinates $\vect{V} = (U,W,V) = (\dot{X}, \dot{Y}, \dot{Z})$, where
the origin lies near the systemic velocity.  As a rule, we can measure
the first two components of $(X,Y,Z)$ and the third component of
$(U,W,V)$, while the other components are effectively unconstrained.

Consequently, if tidal material is detected with a specific position
and line-of-sight velocity, we can only say that the phase-space
region it occupies has finite extent in some directions and infinite
extent in others.  For example, consider a phase-space region
\begin{equation}
  \tilde{\vect{Q}} =
    \{ (X,Y,Z;U,W,V) \in \mathsf{R}^6 \,|\,
       X \in X_Q \pm R_Q, \, Y \in Y_Q \pm R_Q, \, V \in V_Q \pm S_Q \} \, .
  \label{voxel_definition}
\end{equation}
Here $(X_Q, Y_Q)$ specifies a point on the plane of the sky, $R_Q$ is
a radius, $V_Q$ specifies a line-of-sight velocity, and $S_Q$ is a
velocity range.  Such a region $\tilde{\vect{Q}}$ might, for
example, correspond to a voxel in a H{\footnotesize{I}} data-cube.

Simulation coordinates $\vect{r}$ and $\vect{v}$ may be projected onto
the observational plane via
\begin{equation}
  \vect{R} = \mathcal{L} \mathbfss{M} \vect{r} + \vect{R}_\mathrm{c} \, ,
  \qquad
  \vect{V} = \mathcal{V} \mathbfss{M} \vect{v} + \vect{V}_\mathrm{c} \, .
  \label{obs_proj}
\end{equation}
Here the rotation matrix $\mathbfss{M} = \mathbfss{M}(\theta_\alpha)$,
which depends on the Euler angles, rotates simulation coordinates to
observation coordinates.  Assuming the simulation coordinates are
dimensionless, the scale factors $\mathcal{L}$ and $\mathcal{V}$ will
have units of length and velocity, respectively.  Finally,
$\vect{R}_\mathrm{c} = (X_\mathrm{c}, Y_\mathrm{c}, 0)$ and
$\vect{V}_\mathrm{c} = (0,0,V_\mathrm{c})$ specify offsets in the
plane of the sky and the line-of-sight velocity, respectively.  These
offsets vanish if the origins of $\vect{R}$ and $\vect{V}$ coincide
exactly with the system's centre of mass position and velocity, but
this is hard to insure in practice.

It is convenient to define an operator $\mathcal{P}: \vect{q} \mapsto
\vect{Q}$ which implements the mapping from simulation to observation:
\begin{equation}
  \mathcal{P}: (\vect{r}, \vect{v}) \mapsto
    (\mathcal{L} \mathbfss{M} \vect{r} + \vect{R}_\mathrm{c},
     \mathcal{V} \mathbfss{M} \vect{v} + \vect{V}_\mathrm{c}) \, .
\end{equation}
The observational version of (\ref{omega_def}) is then
\begin{equation}
  \Omega_{d}(\vect{s}; \tilde{\vect{Q}}, t,
              \theta_\alpha, \mathcal{L}, \mathcal{V},
              \vect{R}_\mathrm{c}, \vect{V}_\mathrm{c}) \doteq
  \sum\nolimits_{\textstyle \mathcal{P}\vect{q}_i(t) \in \tilde{\vect{Q}}} \,
    \xi_i \delta^2(\vect{s} - \vect{s}_i) \, ,
\end{equation}
where $\mathcal{P}$ depends on the parameters $\theta_\alpha$,
$\mathcal{L}$, $\mathcal{V}$, $\vect{R}_\mathrm{c}$, and
$\vect{V}_\mathrm{c}$; these parameters are involved because the
region $\tilde{\vect{Q}}$ is specified observationally.
Observational versions of $\overline{\Omega}_{d}$ and
$\Omega^*_{d}$ can be defined analogously and likewise depend on
these parameters.

\subsection{Proof of concept tests}
\label{proof_of_concept}

Since observations can constrain only three out of six phase-space
dimensions, it's not clear if the algorithm described above can
yield well-determined disc spins.  One way to find out is to
apply the algorithm to data generated using simulated galaxy
encounters; if the algorithm reconstructs initial disc spins
from simulated observational data then it passes the test.  To make
the test as clean as possible, the simulated encounters are run using
test-particle discs embedded in a spherical self-consistent models,
and the \textit{same} mass model is used in the Identikit simulation
to compute $g_{d}(\vect{r}, \vect{v}, \vect{s}, \xi; t)$.
This side-steps, for now, two potential complications: (1) the
difference between test-particle and self-consistent tidal responses,
and (2) any mismatch between the mass model used for the simulated
data and the one used for the Identikit computation.

\textok{This test uses the same composite mass model adopted in BH09,
with a bulge/disc/halo mass ratio of 1:3:16.  The bulge has a
\citet{H90} profile, the disc has an exponential surface-density
profile, and the halo has a \citet*{NFW96} profile and is tapered
smoothly at $\sim 12$ disc scale lengths.}  As in \textok{BH09}, the
calculations are carried out in units with $G = 1$; in these units the
rotation period at $3$ disc scale lengths is $t_\mathrm{rot}
\simeq 1.23$.  Two discs with spins $(i_{1},\omega_{1}) = (30,0)$ and
$(i_{2},\omega_{2}) = (135,0)$ are placed on approaching parabolic
orbits, reaching periapsis at time $t = 0$.  By time $t = 1$, disc
${1}$ has produced a well-developed tidal tail.  For simplicity, the
observation operator $\mathcal{P}$ is replaced with the identity
operator, so that $\vect{q} = \vect{Q}$.  The simulated orbit lies in
the $(X,Y)$ plane, while the system is viewed along the orbital axis
$Z$.

\begin{figure}
\begin{center}
\includegraphics[clip=true,width=0.55\columnwidth,angle=-90]{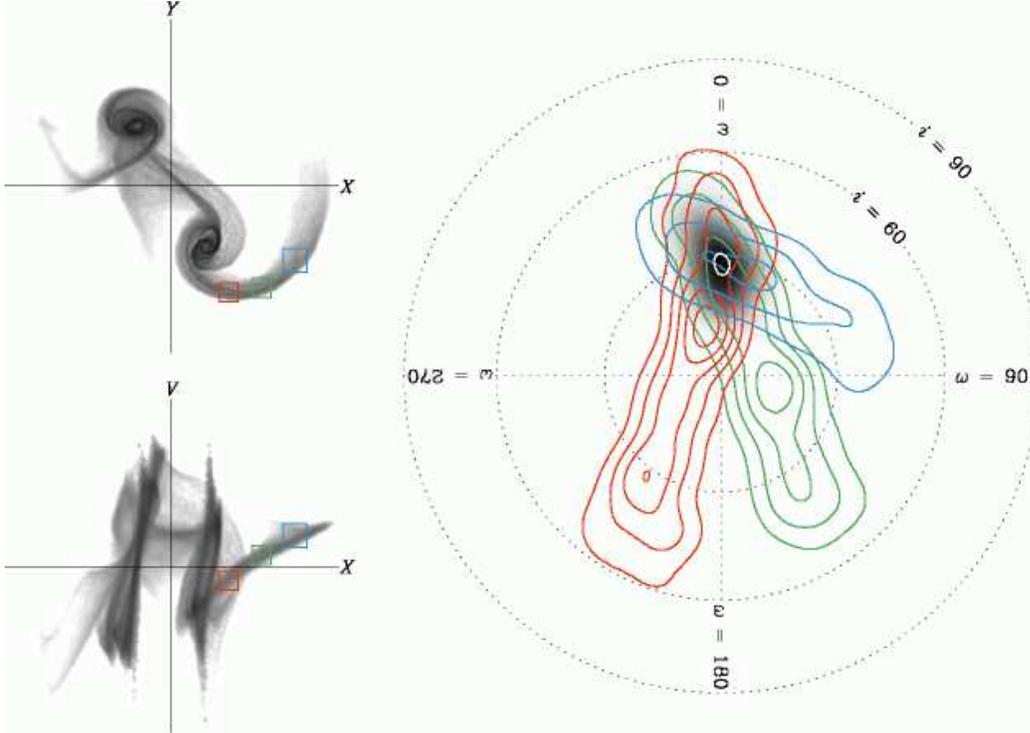}
\caption{A proof-of-concept test of the Identikit~2 algorithm.  On the
left are $(X,Y)$ and $(X,V)$ projections of an encounter between two
disc galaxies with spins $(i_{1},\omega_{1}) = (30,0)$ and
$(i_{2},\omega_{2}) = (135,0)$; the system is viewed along the orbital
axis.  The small colored boxes \textok{show} three phase-space regions
$\tilde{\vect{Q}}_j$ distributed along the tidal tail produced by
galaxy~{1}.  On the right \textok{is an equal-area projection of one
hemisphere of the spin sphere}; lines of constant inclination $i$ and
argument $\omega$ are labeled.  \textok{The contour plots show the
functions $\overline{\Omega}_{1}(\vect{s};
\tilde{\vect{Q}}_j, t)$; their colors correspond to the regions defined
on the left.}  The grey-scale \textok{image shows the product
function} $\Omega^*_{1}(\vect{s}; t)$; the single white contour is set
at $95$\% of the peak value, and neatly encloses the actual disc spin
$(i_{1},\omega_{1}) = (30,0)$.  \label{example1}}
\end{center}
\end{figure}

As the left side of Fig.~\ref{example1} shows, three phase-space
regions $\tilde{\vect{Q}}_j$ are defined along the tidal tail of disc
${1}$.  These regions trace the run of tail velocity $V$ as well as
tail position $(X,Y)$, but -- as (\ref{voxel_definition}) implies --
have infinite extent in $Z$ and in $(U,W)$.  The right side of
Fig.~\ref{example1} shows contours of the three functions
$\overline{\Omega}_{1}(\vect{s}; \tilde{\vect{Q}}_j, t)$, evaluated at
time $t = 1$.  All three regions $\tilde{\vect{Q}}_j$ map to roughly
linear features on the spin sphere; each region can be populated by a
range of disc spins.  However, the three sets of contours overlap in
one place on the spin sphere, and the product function
$\Omega^*_{1}(\vect{s}, t)$ has a single peak very near the true disc
spin $(i_{1},\omega_{1}) = (30,0)$.  The white contour shows where
$\Omega^*_{1}(\vect{s}, t)$ equals $95$\% of its peak value.  This
contour, which neatly encloses the true $(i_{1},\omega_{1})$,
illustrates the uncertainty in the derived spin.

\begin{figure}
\begin{center}
\includegraphics[clip=true,width=0.55\columnwidth,angle=-90]{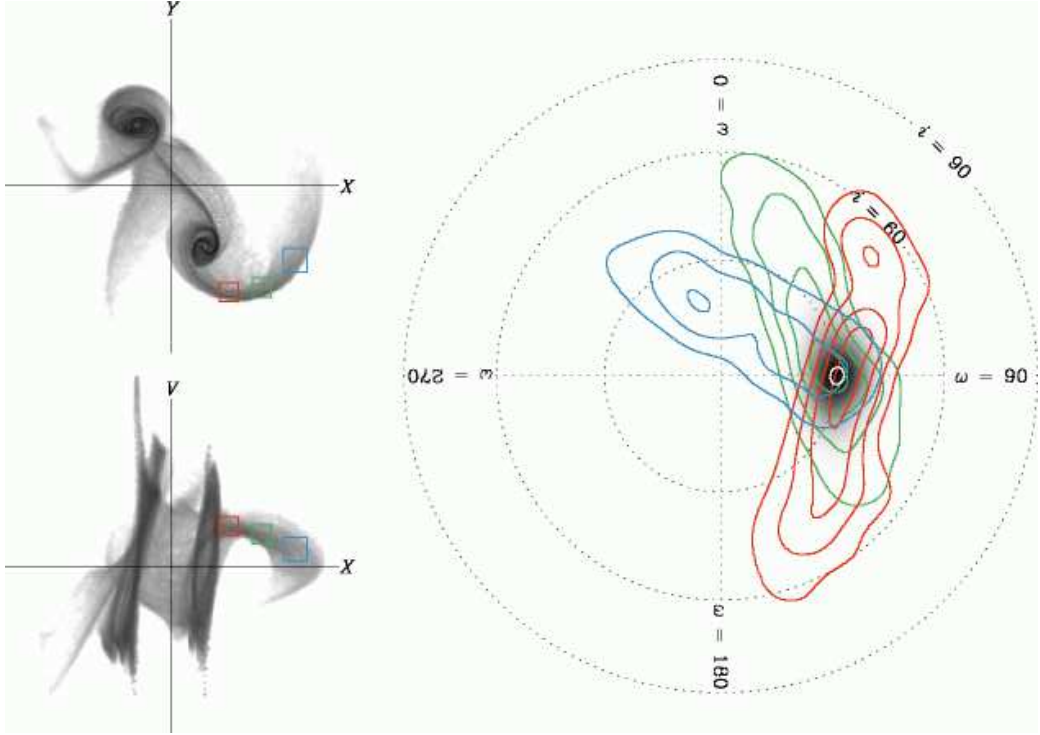}
\caption{Another test of the algorithm.  Here, disc~{1} has spin
$(i_{1},\omega_{1}) = (30,90)$.  The three colored boxes have the same
locations on the $(X,Y)$ plane as in Fig.~\ref{example1}, but have
been shifted in $V$ to track the run of velocities along the tail.  As
a result, the function $\Omega^*_{1}$ now peaks very close to
$(i,\omega) = (30,90)$.  This illustrates the key role velocity
information plays in constraining encounter geometry.
\label{example2}}
\end{center}
\end{figure}

A simple variation of this experiment is shown in Fig.~\ref{example2},
where the argument to periapsis of disc~{1} has been changed to
$\omega_{1} = 90$.  This has little effect on the resulting tail's
morphology, but yields a different run of line-of-sight velocities as
shown on the bottom left of the figure.  The three regions defined
here have the same $(X,Y)$ coordinates as in the previous case; only
their $V$ coordinates have been modified.  In response, the peak of
the product function $\Omega^*_{1}(\vect{s}, t)$ shifts to very near
the actual spin $(i_{1},\omega_{1}) = (30,90)$.  Again, the white
contour at $95$\% of the peak shows how well the spin is constrained.

These simple tests demonstrate several key points.  First, a
\textit{single} phase-space region $\tilde{\vect{Q}}$ defined
following (\ref{voxel_definition}) \textit{doesn't} provide enough
information to uniquely determine initial disc spin; the various
$\overline{\Omega}_{1}(\vect{s}; \tilde{\vect{Q}}_j, t)$ functions
contoured in Figs.~\ref{example1} and~\ref{example2} each admit a
range of possible solutions.  Second, two or more regions
$\tilde{\vect{Q}}_j$ tracing a given tidal structure \textit{can}, at
least in theory, accurately determine initial disc spin.  For best
results the regions used should sample different parts of the tidal
feature originating from a given disc, so the resulting
$\overline{\Omega}_{d}(\vect{s}; \tilde{\vect{Q}}_j, t)$ functions are
not closely aligned with one another.  Third, line-of-sight velocity
information is necessary; morphological constraints alone can't
distinguish between the two cases presented in this section.

\section{Searching Parameter Space}

The algorithm described above yields disc spins assuming that all
other model parameters are \textit{already} known.  This capability
may be useful in examining variations on a known solution, but a more
comprehensive methodology is required to model real systems from
scratch.  Some \textok{technique for} searching parameter space or
solving for the other parameters is needed; this section develops a
possible approach.

\subsection{Measuring quality of fit}
\label{measuring_fit_quality}

A numerical measure of the quality of a solution is needed to
implement an automated search of parameter space.  The function
$\Omega^*_{d}(\vect{s}; t, \dots)$, where $(t, \dots)$ represents all
orbit and viewing parameters, may be useful in this role.  If
$\Omega^*_{d}$ vanishes for \textit{all} disc spins $\vect{s}$, then
either:
\begin{enumerate}
\renewcommand{\theenumi}{(\arabic{enumi})}
\item
The regions $\tilde{\vect{Q}}_j$ do not all originate from a
single disc with a unique spin vector.
\item
The adopted orbit parameters $(p, \mu, e)$ or viewing parameters $(t,
\theta_\alpha, \mathcal{L}, \mathcal{V}, \vect{R}_\mathrm{c},
V_\mathrm{c})$ are too inaccurate to yield a solution.
\end{enumerate}
The first interpretation may be correct if the disc was initially
warped or if some of the regions are attributed to the wrong galaxy.
In most cases, however, the second interpretation may be taken as a
working hypothesis.

Extending this line of thought, it seems plausible to prefer model
solutions which yield higher peak values of $\Omega^*_{d}$.  Let
\begin{equation}
  \Lambda_{d}(t, \dots) =
    \max\nolimits_{\textstyle \,\vect{s}} \,
      \Omega^*_{d}(\vect{s}; t, \dots) \, ,
\end{equation}
be the maximum value of $\Omega^*_{d}$ obtained for any $\vect{s} \in
\mathsf{S}^2$.  The peak value basically measures the degree to which
the functions $\overline{\Omega}_{d}$ for different regions overlap
with one another; a solution in which the ridge-lines of these
functions intersect is probably better than one in which their
outskirts barely touch.

Just two regions, suitably chosen, suffice to constrain a disc's spin
if all other parameters are known.  However, for $n_\mathrm{reg} = 2$
regions the peak function $\Lambda_{d}(t, \dots)$ will be fairly
insensitive to the viewing and orbit parameters.  An inaccurate choice
for these parameters will probably shift the locus of the overlap
between the images of $\tilde{\vect{Q}}_1$ and $\tilde{\vect{Q}}_2$,
but won't appreciably reduce $\Lambda_{d}$ unless the error is quite
large.  To make $\Lambda_{d}$ a useful metric, at least
$n_\mathrm{reg} = 3$ regions should be used to trace each disc's tidal
features, and it seems likely that more regions will yield stronger
constraints on the collision parameters.

Finally, a plausible model of a galactic collision must be able to
account for the tidal features of \textit{both} galaxies using the
\textit{same} set of orbit and viewing parameters.  Suppose functions
$\Omega^*_{1}(\vect{s}; t, \dots)$ and
$\Omega^*_{2}(\vect{s}; t, \dots)$ have been defined using two
different sets of regions, each tracing the tidal features of one of
the galaxies.  By analogy with (\ref{omegastar_def}), let
\begin{equation}
  \Lambda(t, \dots) =
    \Lambda_{1}(t, \dots) \, \Lambda_{2}(t, \dots) \, ;
  \label{omega_hat_combined}
\end{equation}
this product, which depends on all orbit and viewing parameters,
provides an estimate of the overall quality of a solution.

\subsection{Centre `locking'}

In many cases, a subset of the viewing parameters may be determined
straightforwardly.  Even violently disturbed galaxies often have
well-defined nuclei which can be accurately located, and a plausible
simulation of an interacting pair should reproduce the positions of
these nuclei.  As BH09 noted, when modeling a system in which two
galaxies appear well-separated on the plane of the sky, it's helpful
to `lock' the centres of the models.  Locking derives the rotation
about the viewing axis $\theta_Z$, length scale $\mathcal{L}$, and
offset $\vect{R}_\mathrm{c}$ by requiring the centres of the models to
coincide with the nuclei of the real galaxies.  These four viewing
parameters, in effect, become functions of the viewing direction
$(\theta_X, \theta_Y)$, time since periapsis $t$, and orbital
parameters $(p, \mu, e)$.  Locking is completely independent of disc
orientation, so it can be applied \textit{before} attempting to solve
for disc spins.

\subsection{Scanning over viewing direction}

If the pair of galaxies to be modeled exhibit distinct and
well-separated nuclei, it's straightforward to perform a systematic
search of all viewing directions.  Suppose for the moment that trial
values have been chosen for the orbit parameters $(p, \mu, e)$, time
since periapsis $t$, and velocity scale $\mathcal{V}$ and offset
$V_\mathrm{c}$.  The basic idea is to iterate over a lattice of
possible viewing directions parametrized by the Euler angles
$(\theta_X, \theta_Y)$.  For each viewing direction, centre locking
determines the remaining viewing parameters, and a trial solution for
disc spins is obtained using the algorithm in \S~\ref{deriving_spins};
the best viewing direction is the one which yields the largest value
of $\Lambda$.  In effect, this prescription surveys a ten-parameter
subset of the full parameter space (Fig.~\ref{parameters}) at the cost
of a blind search of only two parameters.

In the present implementation, the lattice of viewing directions is
generated by starting with an icosahedron inscribed within a unit
sphere, and replacing every one of its $20$ triangular faces with four
equilateral triangles, each with one quarter of the original
triangle's area.  The new vertexes introduced are projected out to
unit radius, producing a solid with $n_\mathrm{face} = 80$ triangular
faces approximating a sphere.  This process can be iterated, producing
a sequence of ever-better approximations with $n_\mathrm{face} = 320$,
$1280$, $5120$, \dots faces.

The midpoints of these faces, which cover the sphere in a nearly
uniform fashion, define a lattice of viewing directions.  Each face,
indexed by $k$ (where $k = 1, \dots, n_\mathrm{face}$), is associated
with a viewing direction $(\theta_{X,k}, \theta_{Y,k})$.  Locking
determines corresponding values for the angle $\theta_{Z,k}$, scale
$\mathcal{L}_k$, and offset $\vect{R}_{\mathrm{c},k} =
(X_{\mathrm{c},k}, Y_{\mathrm{c},k}, 0)$.  Solving for disc spins
yields the peak values
\begin{equation}
  \Lambda_{{d},k}(t,\mathcal{V},V_\mathrm{c},p,\mu,e) =
    \Lambda_{d}(t, \theta_{\alpha,k}, \mathcal{L}_k,
                            \mathcal{V},\vect{R}_{\mathrm{c},k},
                            V_\mathrm{c},p,\mu,e) \, .
\end{equation}
In addition to finding the viewing direction $k$ which maximizes the
product $\Lambda_k = \Lambda_{{1},k}
\Lambda_{{2},k}$, the algorithm tabulates values of
$\Lambda_{{d},k}$ for all faces.

\section{Testing Parameter Search}
\label{test_param_search}

The approach just outlined only works if the product $\Lambda(t,
\dots)$ can reliably identify good solutions.  To establish this, the
next test treats the viewing angles $\theta_\alpha$, length scale
$\mathcal{L}$ and position offset $\vect{R}_\mathrm{c}$ as unknowns,
in addition to the disc spins $(i_{d}, \omega_{d})$.  Conversely, the
orbit parameters $(p, \mu, e)$, time since periapsis $t$, and velocity
scale $\mathcal{V}$ and offset $V_\mathrm{c}$ have known values.
While this test therefore doesn't search the full parameter space, it
does survey a non-trivial subset; in particular, this test aims to
determine if Identikit~2 can reconstruct the encounter \textit{and}
viewing geometry for well-separated pairs of interacting galaxies.

\begin{figure}
\begin{center}
\includegraphics[clip=true,width=6.7333in]{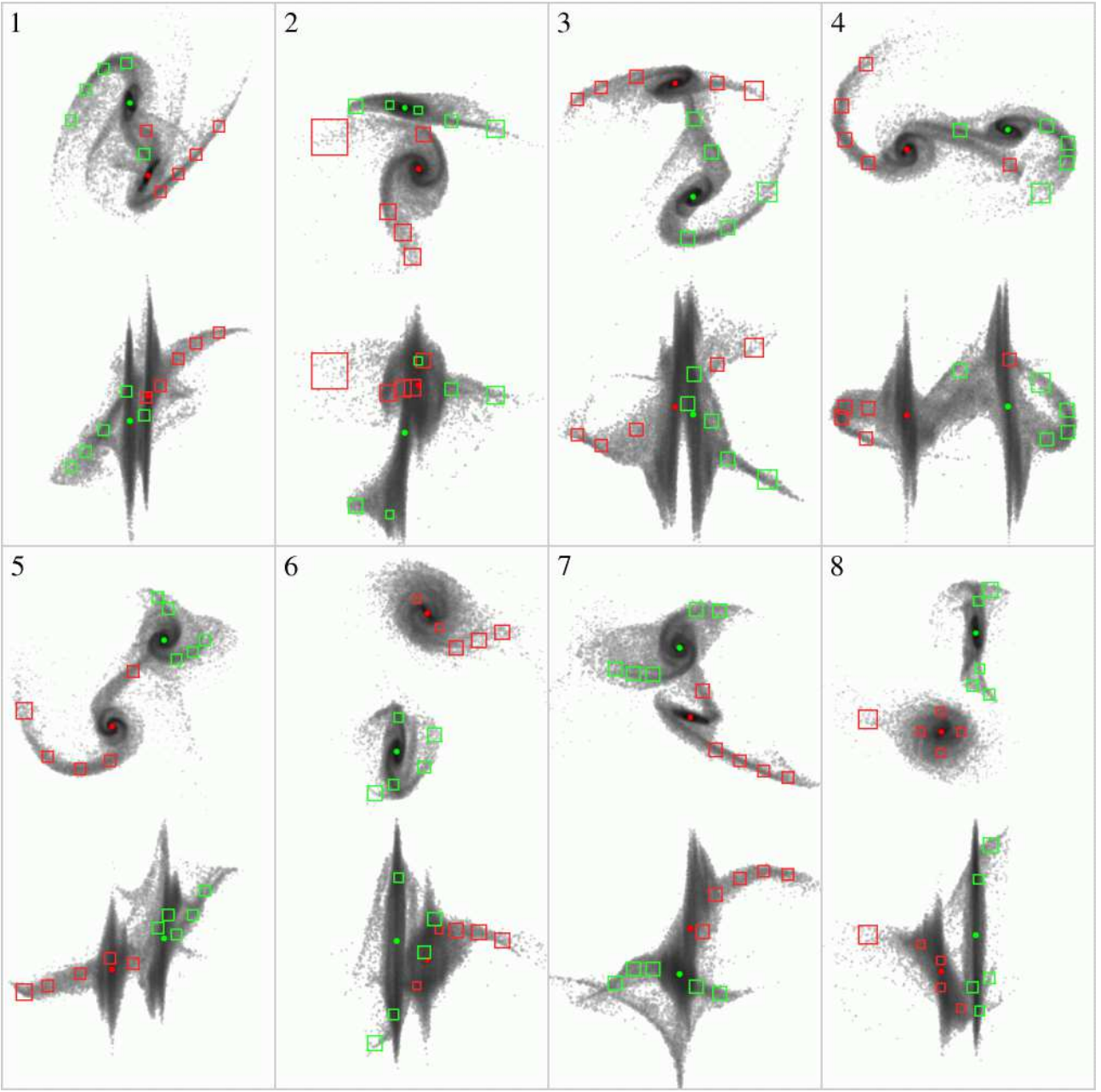}
\caption{$(X,Y)$ and $(X,V)$ projections of eight self-consistent galaxy
encounters used to test the Identikit~2 algorithm, \textok{represented
by grey-scale images.  The nuclear position and velocity of each
galaxy are shown by red and green dots for galaxy~{1} and galaxy~{2},
respectively.}  Five regions, \textok{color-coded according to the
same convention}, are used to trace the tidal features of each galaxy
disc.  \label{view_box}}
\end{center}
\end{figure}

Fig.~\ref{view_box} presents a sample of equal-mass ($\mu = 1$)
initially-parabolic ($e = 1$) galaxy encounters with random disc spins
and viewing angles.  Out of the $36$ self-consistent encounters BH09
used to test Identikit~1, these are the eight with the largest
periapsis separations; they include an equal mix of direct and
retrograde discs.  All eight are `observed' at $t_\mathrm{true} =
1$, one model time unit after first periapsis.  At this time, these
pairs all have well-separated centres, insuring that centre locking
will work effectively.

Following BH09, the Identikit calculations use a spherically-averaged
version of the \textit{same} mass model employed in the
self-consistent simulations.  Since the range of actual periapsis
separations $p_\mathrm{true}$ represented in this sample is fairly
small, all eight were matched using a single Identikit simulation with
$p$ comparable to the mean separation.  This simulation was evolved to
$t = 1$, exactly matching the actual time.

Five regions, shown in Fig.~\ref{view_box}, were used to trace the the
tidal features of each disc.  For the most part, these regions simply
follow the morphology and kinematics of the tidal structures seen in
$(X,Y)$, $(X,V)$, and $(V,Y)$ projections of the self-consistent
simulations.  In some cases, however, these projections don't fully
delineate the information contained in a $(X,Y,V)$ data-cube.  An
interactive routine was therefore used to evaluate the amount of tidal
material in each region as its parameters were adjusted; this insured
that the regions actually contain significant amounts of material even
when the projected views were ambiguous.  Of course, the same approach
could be used when working with observational data.

\begin{figure}
\begin{center}
\includegraphics[clip=true,width=6.7333in]{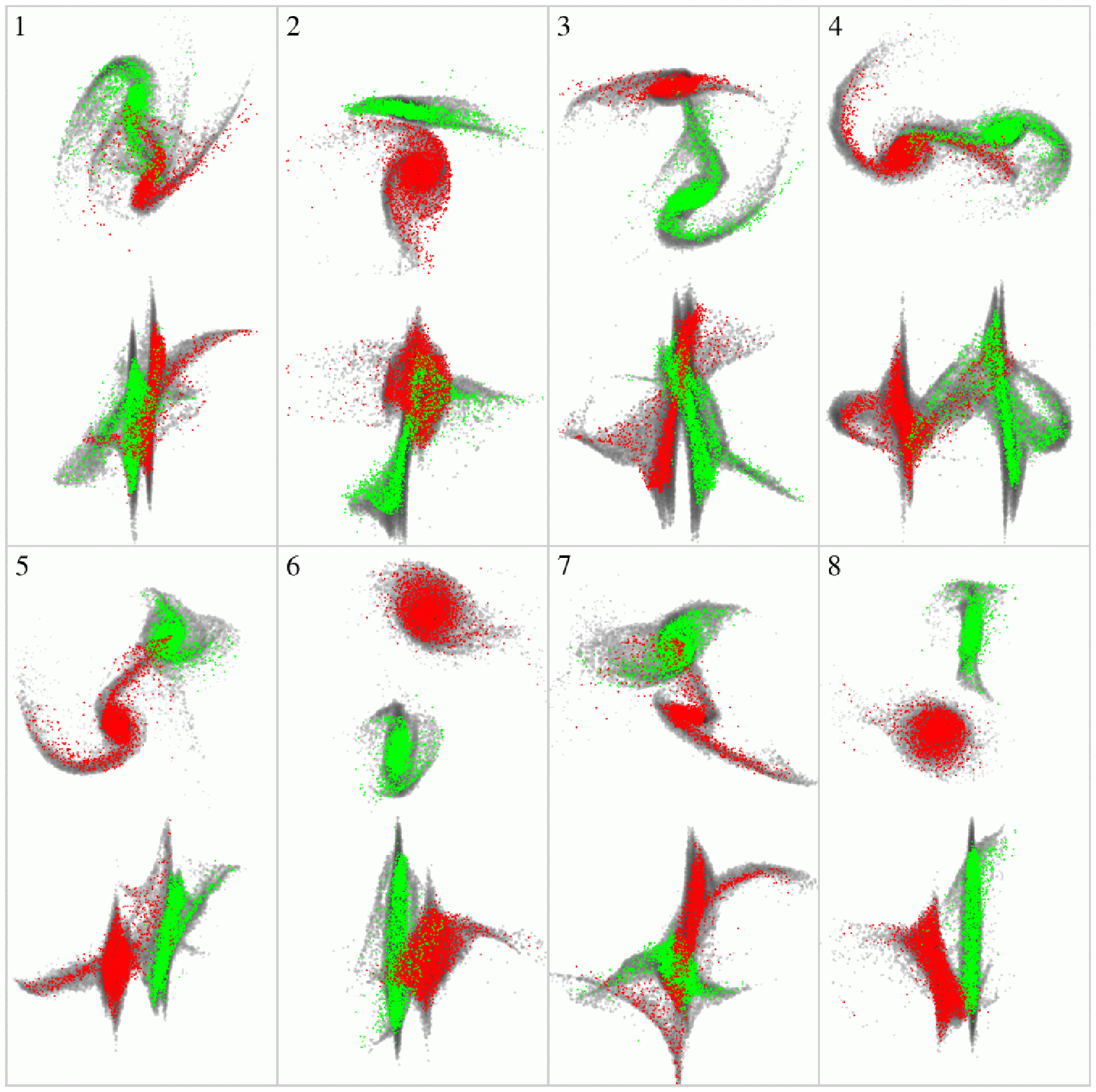}
\caption{$(X,Y)$ and $(X,V)$ projections of Identikit~2 models for
the eight galaxy encounters. \textok{Red and green points show
solutions for disc~{1} and disc~{2}, respectively, over-plotted on
grey-scale images of the self-consistent simulations.}
\label{view_pnt}}
\end{center}
\end{figure}

Once these regions were defined, a preliminary model for each system
was found by scanning over a lattice of $n_\mathrm{face} = 320$ to
$1280$ viewing directions.  In two cases (encounters~2 and~5), the
algorithm failed these preliminary tests.  The nature of these
failures will be discussed shortly, but fairly minor adjustments to
the regions sampling just one of the two discs were enough to resolve
both cases.  With these adjustments made, final models for all eight
systems were obtained by scanning $n_\mathrm{face} = 5120$ viewing
directions.  Each of these final models took about one hour of
computing time on a $3 \,\mathrm{Ghz}$ Intel processor; this time
could be substantially reduced by optimizing and parallelizing the
code.  Fig.~\ref{view_pnt} shows these eight models (points),
overplotted on the self-consistent simulations (grey-scale images).
The algorithm successfully reproduced the morphologies and kinematics
of all eight systems; by the standards of BH09, these are all `good'
matches.

\begin{figure}
\begin{center}
\includegraphics[clip=true,width=6.7333in]{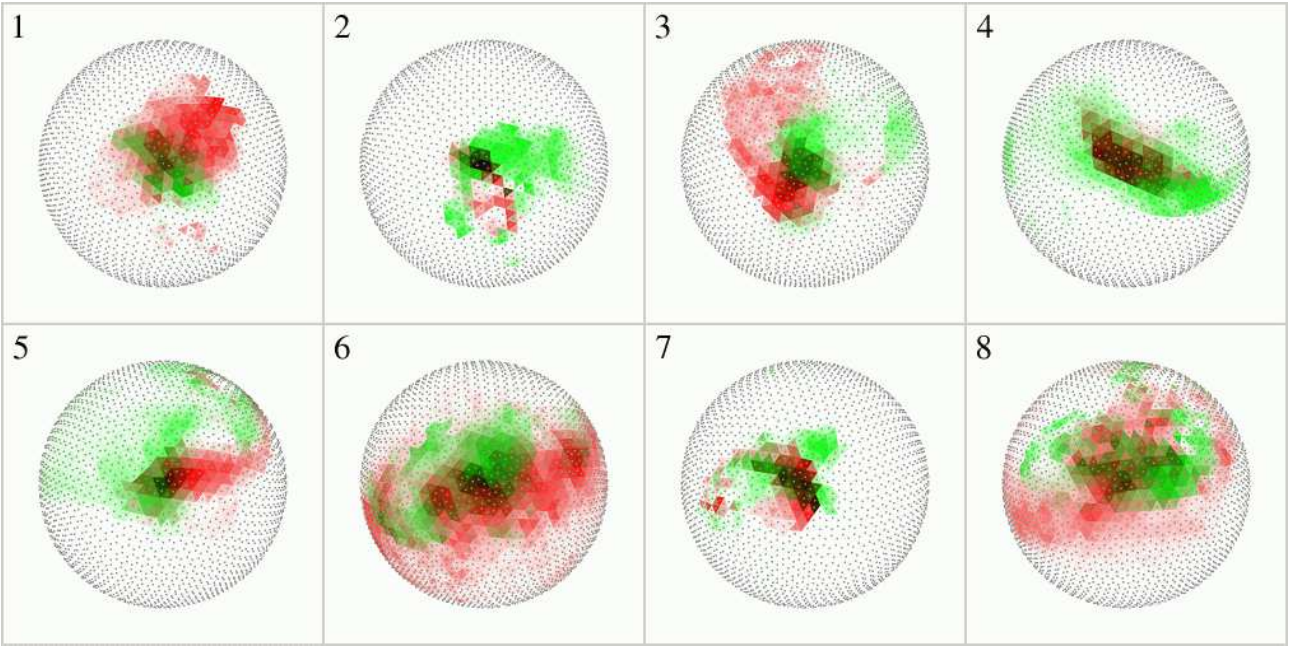}
\caption{Perspective views of the viewing direction lattice with color
representations of $\Lambda_{d}$, \textok{using red for disc~{1} and
green for disc~{2}}.  \label{view_ball}}
\end{center}
\end{figure}

The algorithm selects viewing direction by maximizing the product
$\Lambda_{{1},k} \Lambda_{{2},k}$, but how do the constraints provided
by the two discs work together?  Fig.~\ref{view_ball} shows how
$\Lambda_{d}$ varies as a function of viewing direction.  Each panel
shows a perspective view of the viewing direction lattice, with the
viewing direction selected by the algorithm situated dead centre.  The
color of each triangular face shows $\Lambda_{{1},k}$ and
$\Lambda_{{2},k}$, using red for disc~{1} and green for disc~{2}.  The
eight models yield a wide range of `landscapes' on this spherical
lattice.  Somewhat predictably, models~6 and~8, in which both discs
display only subtle tidal disturbances, yield diffuse $\Lambda_{d}$
functions with multiple local maxima.  More pronounced tidal
disturbances, such as the classic `bridge and tail' structures seen in
both discs of models~1 and~4, and in one disc each in models~2, 3, 5,
and~7, generally yield more localized $\Lambda_{d}$ functions; while
multiple maxima may still appear, they are closer together.  In
models~2, 3 and~5, the individual $\Lambda_{d}$ functions appear
nearly disjoint, only overlapping for a small range of viewing
directions.  Model~4, in contrast, produces $\Lambda_{d}$ functions
which are roughly aligned with each other, although one function is
considerably more diffuse than the other.  Clearly, the intrinsic
tidal response, viewing direction, and strategy used to pick regions
are all factors in determining the landscape of the resulting
$\Lambda_{d}$ functions; further experimentation with a larger set of
encounters may help to tease these factors apart.

\begin{figure}
\begin{center}
\includegraphics[clip=true,width=0.5\columnwidth]{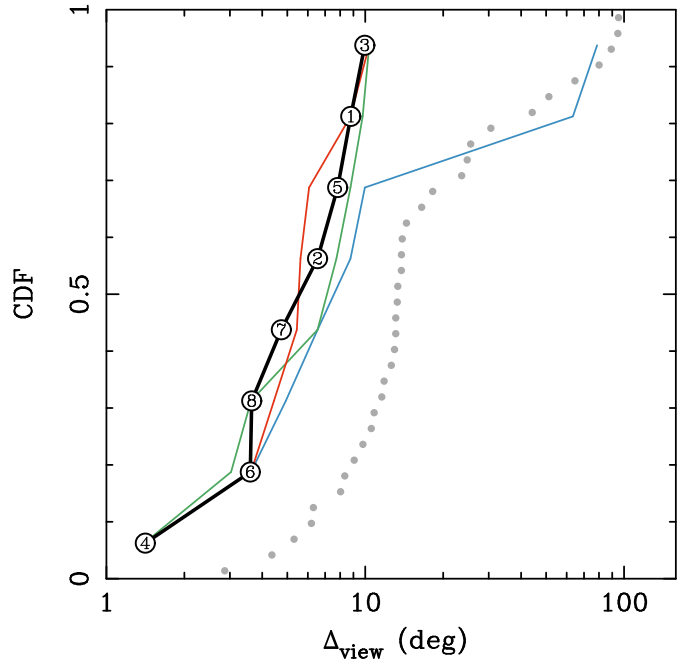}
\caption{Cumulative distribution functions for the error in viewing
direction $\Delta_\mathrm{view}$.  The numbered points show results
for the eight final models.  Colored lines show variants of the
algorithm: \textok{red shows results for equal particle weights
($\xi_i = 1$), blue for unnormalized peaks ($K(\tilde{\vect{q}}) =
1$), and green for both together ($\xi_i = K(\tilde{\vect{q}}) = 1$)}.
The light grey dots show results from BH09.
\label{cdf_view}}
\end{center}
\end{figure}

In practice, the method used to select viewing direction works quite
well for this set of galaxy encounters.  Fig.~\ref{cdf_view} shows the
cumulative distribution of angular errors in viewing direction
$\Delta_\mathrm{view}$.  Numbered points represent the eight
Identikit~2 models from Fig.~\ref{view_pnt}, while the light grey dots
show the set of 36 Identikit~1 models from BH09.  The median error in
viewing direction for the Identikit~2 models is $\Delta_\mathrm{view}
\simeq 5.6\degr$, which is less than half the median error for the
Identikit~1 models.  Moreover, \textit{all} of the new models are good
fits; the tail of poor solutions represented by the grey dots
extending to the upper right is absent.  Of course, this comparison is
not entirely fair; BH09 also fit for the time $t$, periapsis
separation $p$, and velocity scale $\mathcal{V}$, while here these
parameter values are pre-determined.  It's nonetheless remarkable that
models~6 and~8, which would probably be quite tricky to model by hand,
are among the most accurate of the eight models plotted.  Inspection
of Fig.~\ref{view_ball} suggests that the best solutions arise when
the peaks of the two $\Lambda_{d}$ functions fall fairly near each
other, while somewhat less accurate results are obtained when the
selected viewing direction is a compromise between separate peaks, as
in models~2, 3 and~5.

\begin{figure}
\begin{center}
\includegraphics[clip=true,width=0.5\columnwidth]{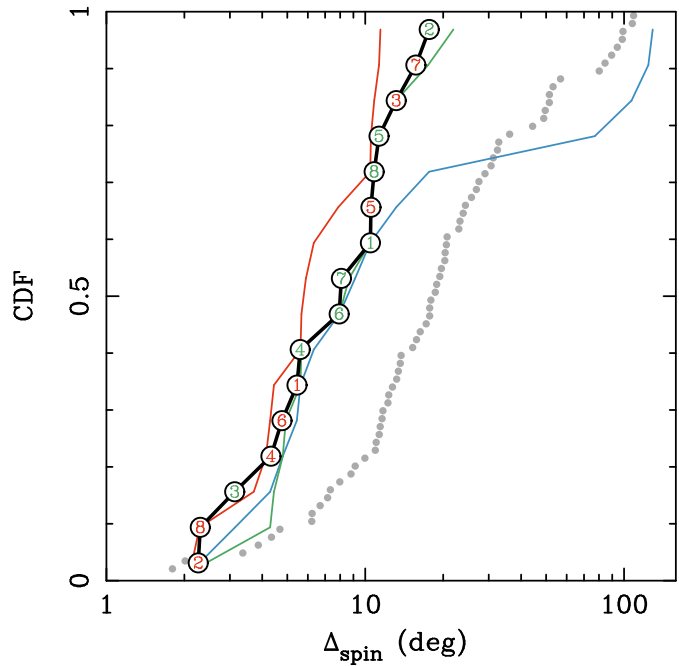}
\caption{Cumulative distribution functions for the error in spin
direction $\Delta_\mathrm{spin}$.  The numbered points show results
for the eight final models, with \textok{red for disc~{1} and green
for disc~{2}}.  Colored lines show results for variants of the
algorithm \textok{as in Fig.~\ref{cdf_view}}.  The light grey dots
show results from BH09; one dot, off-scale to the left, is not plotted
\label{cdf_spin}}
\end{center}
\end{figure}

Fig.~\ref{cdf_spin} plots the cumulative distribution of angular
errors in disc spin direction $\Delta_\mathrm{spin}$.  Here the
numbered points represent the sixteen discs from Fig.~\ref{view_pnt},
color coded as in that figure.  Given that the viewing directions for
the eight Identikit~2 models are all determined with an error of $\sim
10\degr$ or less, it's not too surprising that the spins of the
individual discs are also well determined.  The median error for this
sample is $\Delta_\mathrm{spin} \simeq 8.0\degr$, again less than half
of the analogous error for the BH09 sample (grey dots).  In
conjunction with Fig.~\ref{view_pnt}, this plot reveals some
interesting systematics.  As a group, the discs with accurate spin
directions tend to be face-on, while those with large errors tend to
be edge-on.  This is neatly illustrated by model~2, where galaxy~{1}
(red) is fairly face-on and has a spin error $\Delta_\mathrm{spin} =
2.3\degr$, while galaxy~{2} (green) is nearly edge-on and has
$\Delta_\mathrm{spin} = 17.7\degr$.  \textok{The same} pattern is also
seen in model~8 ($2.3\degr$ vs. $10.8\degr$) and to a smaller degree
in model~6 ($4.8\degr$ vs. $7.9\degr$).  \textok{This trend may arise
because face-on discs present truly two-dimensional velocity fields,
while edge-on discs collapse the two spatial dimensions into one; in
effect, the former provide more information.}  On the other hand, the
error in spin direction $\Delta_\mathrm{spin}$ does not appear to
correlate with disc inclination $i$.  The four discs at the bottom of
Fig.~\ref{cdf_spin}, which have the most accurately determined spins,
are an equal mix of direct ($i < 90\degr$) and retrograde ($i >
90\degr$) passages; the four at the top include three retrograde
passages, but the three discs just below this group are all direct.
This is somewhat unexpected since direct passages produce pronounced
tidal bridges and tails, which would seem to offer better constraints
on model solutions; it is, however, consistent with the very accurate
results already noted for the purely retrograde encounters~6 and~8.

The colored lines in Figs.~\ref{cdf_view} and~\ref{cdf_spin} show
results for different versions of the disc spin algorithm
(\S~\ref{deriving_spins}).  The red curves in these figures were
obtained by calculating $\Omega_{d}$ (\ref{omega_def}) with all
particles weighted equally (in effect, setting $\xi_i = 1$).  The blue
curves result from computing $\overline{\Omega}_{d}$
(\ref{omegabar_def}) without normalizing the peak value to unity
(setting $K(\tilde{\vect{q}}) = 1$).  Finally, the green curves
combine both options (setting $\xi_i = K(\tilde{\vect{q}}) = 1$).
For the most part, these different variants produce similar results.
Equal weighting, which effectively overrepresents the outer regions of
the initial discs, may yield a modest increase in spin direction
accuracy.  Conversely, doing without normalization produces less
accurate models, although in most cases the differences are slight
-- the exceptions will be discussed next.

\subsection{Failed models}

Fig.~\ref{cdf_view} shows that two of the models constructed without
normalizing $\overline{\Omega}_{d}$ (blue curve) are clearly
\textit{very} poor solutions.  These models have errors in viewing
direction of $\Delta_\mathrm{view} = 63\degr$ and $79\degr$, which
induce comparable errors in spin direction, as shown in
Fig.~\ref{cdf_spin}.  A similar failure, with $\Delta_\mathrm{view} =
72\degr$, was seen in a preliminary model for encounter~5.  In all of
these cases, the algorithm selected a viewing direction roughly
parallel to the line between the two galaxies, so that the centres of
the models appear close in projection.  Because the model centres are
forced to coincide with the nuclei of the self-consistent galaxies,
these viewing directions yield very large values for the scale factor
$\mathcal{L}$.  As a result, the transformation from simulation to
observation coordinates (\ref{obs_proj}) splatters particles into all
of the regions $\tilde{\vect{Q}}_j$ used to constrain model
solutions.  Since both model galaxies are scaled by the same
$\mathcal{L}$, both $\Lambda_{d}$ functions have local peaks for the
same viewing direction, and their product $\Lambda$ rises higher than
the peak representing the true viewing direction.

This mode of failure is more likely when $\overline{\Omega}_{d}$ is
not normalized because the particles splattered into the
$\tilde{\vect{Q}}_j$ regions often come from the inner parts of
the galaxies and therefore have relatively large weights $\xi_i$.
These high-weight particles help to push the false peak of $\Lambda$
above the one corresponding to the true solution.  If, in addition to
not normalizing $\overline{\Omega}_{d}$, all particles are given equal
weight (green curve in Fig.~\ref{cdf_view}), then viewing directions
which scatter particles from small radii far and wide are less favored
and failure becomes less likely.  However, using equal weights when
the test-particle distribution is actually biased by a factor of $r^2$
seems rather arbitrary.

In practice, failed solutions such as these are easy to recognize
visually, and it would also be straightforward to detect them
automatically.  One way to do so is to place astrophysically motivated
limits on $\mathcal{L}$; for example, a solution which requires discs
an order of magnitude larger than the discs of real galaxies could
presumably be rejected out of hand.  Another way is to detect and
reject solutions which splatter particles far beyond the
\textit{actual} extent of the tidal features.

For completeness, the other failure of the algorithm, which occurred
in the first attempt to model encounter~2, should also be mentioned.
In this encounter, disc~{2} is nearly edge-on, and presents a wide
range of line-of-sight velocities.  The two small regions on either
side of the centre of this disc in Fig.~\ref{view_box} were initially
set to opposite extremes of this velocity range.  But as BH09 showed,
Identikit models with test particles on circular orbits generally
don't reproduce the full velocity width of self-consistent discs.
Consequently, $\Lambda_{{2},k}$ vanished for almost all viewing
directions, and did not overlap at all with $\Lambda_{{1},k}$.
Reducing the range of velocities sampled by these two regions produced
the satisfactory solution shown in Fig.~\ref{view_pnt}.

Finally, it's worth noting that these failures are only partial.  In
every case, an acceptable solution could be recovered by ignoring one
of the two $\Lambda_{d}$ functions and selecting a viewing direction
corresponding to a peak of the other.  With a good match to the
viewing direction selected, the algorithm could then compute good
solutions for the spins of \textit{both} discs.

\section{Discussion}

The problem of finding a model matching the morphology and kinematics
of a pair of colliding galaxies has generally been solved \textok{by}
a process of trial-and-error, informed by physical insight into the
dynamics of tidal interactions.  In this approach, observational data
can't be used to derive initial conditions directly; instead, the
outcome of a model calculation is compared to the observations, and
the initial conditions are modified on the basis of this comparison.
Identikit~2 offers a shortcut: an important component of the initial
conditions -- the initial spin vectors of the interacting discs -- can
be derived directly from the observed morphology and kinematics of the
tidal features.  Moreover, by providing a way to assess the quality of
a solution, the algorithm can be used to search parameter space
automatically.

The Identikit~2 algorithm derives the initial spin vector of a tidally
interacting disc by simultaneously populating a set of phase-space
regions which trace that disc's tidal features.  It may seem odd not
to make any use of the \textit{amount} of material found in each
region, but this `omission' is deliberate.  The tracers commonly used
to study the kinematics of interacting systems don't necessarily obey
a continuity equation; for example, neutral hydrogen may be ionized or
converted to molecular form, so the amount of H{\footnotesize{I}}
found in a given region can't be predicted by purely dynamical models.
On the other hand, even if much of the H{\footnotesize{I}} in a tidal
feature has been converted to \textok{another phase}, the remaining
\textok{H{\footnotesize{I}} may} still provide a useful constraint on
disc spin as long as it has followed a free-fall trajectory.

Since the algorithm does not use a $\chi^2$ statistic to quantify
goodness-of-fit, it's not straightforward to obtain precise confidence
limits on solutions.  Nonetheless, inspection of the function
$\Lambda$ and its constituent factors (Fig.~\ref{view_ball}) offers
some insight into a solution's uniqueness and accuracy.  It may be
worth exploring the landscape of this function in more detail.  For
example, instead of simply maximizing $\Lambda$, the algorithm could
examine solutions around the peak, and look for secondary peaks which
may represent alternate solutions.  A similar examination of the
functions $\Omega^*_{d}$ could likewise reveal uncertainties and
alternate solutions for spin direction.

\subsection{Other parameters}
\label{other_parameters}

The version of the algorithm tested here performs a blind search of
viewing direction, parameterized by $(\theta_X, \theta_Y)$.  It
constrains four other viewing parameters -- the line-of-sight rotation
$\theta_Z$, the length scale $\mathcal{L}$, and two components of the
offset $\vect{R}_\mathrm{c}$ -- from the positions of the galaxy
centres, and computes initial disc spins $(i_{d}, \omega_{d})$ using
the regions tracking each disc.  This accounts for ten parameters out
of the sixteen described in Fig.~\ref{parameters}.  Preliminary
experiments indicate that some of the other parameters can also be
determined by maximizing $\Lambda(t, \dots)$.  For example, the test
in \S~\ref{test_param_search} was repeated varying the time since
periapsis $t$ between $0.5$ and $1.5$; the times maximizing $\Lambda$
clustered around the actual value $t_\mathrm{true} = 1$, with an
r.m.s.~of $0.18$.  Further tests with multiple parameters in play are
necessary; it will be interesting to see if the algorithm can
\textit{simultaneously} fit for the time $t$, periapsis separation
$p$, and velocity scale $\mathcal{V}$ as \textok{well as} BH09 did.

The number of unknowns remaining depends on the type of system to be
modeled as well as the nature of the data available.  Suppose that
accurate systemic velocities for both members of a pair of
well-separated galaxies are available.  By forcing the model nuclei
\textok{to} coincide with their real counterparts in velocity as well
as position, locking can determine the velocity scale $\mathcal{V}$
and offset $V_\mathrm{c}$ in addition to $\theta_Z$, $\mathcal{L}$,
and $\vect{R}_\mathrm{c}$.  Adopting $e = 1$ and using photometry to
estimate the mass ratio $\mu$ leaves just four parameters -- the
viewing direction $(\theta_X, \theta_Y)$, the periapsis separation
$p$, and time since periapsis $t$ -- as unknowns.  In this case a
blind search of these parameters seems reasonable.  However,
sufficiently accurate systemic velocities may be hard to determine;
galactic nuclei have large velocity dispersions, and different tracers
(stars, H${}_\alpha$, H{\footnotesize{I}}, CO) often give results
differing by several tens of km/s.  Only in cases where the systemic
velocities of the nuclei differ by more than the uncertainties is
velocity locking likely to yield useful constraints.

At the other extreme, fully merged systems such as NGC~7252
\citep{S77, HGvGS94, HM95} represent the most difficult class of
objects to model.  The position and systemic velocity of a merger
remnant constrain the offsets $\vect{R}_\mathrm{c}$ and
$V_\mathrm{c}$, but even assuming $e = 1$, a total of \textit{eight}
orbit and viewing parameters (Fig.~\ref{parameters}) remain
indeterminate.  Since the real galaxies have already merged, only
models run past merger need be considered.  Nonetheless, the available
parameter space is still very large, and blindly searching for
possible solutions may not be very rewarding.

Several groups have used genetic algorithms to automate the search for
models of interacting galaxies \citep{W98, TK01}.  In these
algorithms, a population of candidate solutions compete to match the
observational data; the less successful candidates are eliminated, and
the most successful reproduce to replenish the population.  After
enough generations have passed, the population converges toward a
solution matching the observations.  Unlike simple `hill-climbing'
strategies, genetic algorithms are unlikely to be trapped by local
maxima (a familiar \textok{problem with} naive trial-and-error
modeling).  To date, most genetic algorithms have determined the
reproductive fitness of candidate solutions by comparing
low-resolution images of the actual system and candidate pixel by
pixel, with only limited use of velocity data \citep{WD01}.  For such
a comparison to be truly meaningful, the observed material --
typically stars or H{\footnotesize{I}} -- must obey a continuity
equation; as noted above, this may be violated in real systems.
Moreover, as Figs.~\ref{example1} and~\ref{example2} illustrate,
morphology alone is not enough to strongly constrain disc spins;
low-resolution versions of the $(X,Y)$ images in the upper left of
these figures would be almost indistinguishable.

If further tests confirm that maximizing $\Lambda(t, \dots)$ is an
effective way of constraining viewing and orbit parameters, it may be
possible to combine Identikit~2 with a genetic algorithm.  In this
hybrid approach, each member of the candidate population would define
a specific choice of viewing direction, periapsis time and separation,
velocity scale, and possibly orbital eccentricity.  The remaining
viewing parameters would be determined by locking the centres, and
initial spin directions could then be derived directly.  The resulting
$\Lambda$ value could be used to determine the fitness of the
candidate solution.  This approach combines the strengths of both
algorithms.  A genetic algorithm should be able to out-perform a blind
search without getting stuck on local maxima.  Meanwhile, Identikit~2
could efficiently determine disc spins and provide a robust way to
define reproductive fitness which fully includes velocity information
and does not assume continuity.

\subsection{Mass models}
\label{mass_models}

The mass models adopted in Identikit simulations will almost certainly
influence the algorithm's accuracy.  In the tests presented here, the
\textit{same} mass model has been used to construct self-consistent
simulations and their Identikit reproductions.  However, rotation
curves of real disc galaxies exhibit a variety of shapes \citep*{CvG91,
CGH06}; small galaxies often have rising curves, while massive
galaxies may have flat or even falling curves.  This diversity
presumably arises because real galaxies have a range of initial
angular momenta, bulge/disc/halo mass ratios, and assembly histories
\citep*{MMW98}.  Several studies have shown that rotation curve shape
(equivalently, potential well or halo structure) strongly influences
the development of tidal features; in particular, long tidal tails can
be inhibited by sufficiently deep galactic potential wells
\citetext{\citealp*{DMH96}, \citeyear{DMH99}; \citealp{SW99}}.  What
happens if the mass model used in an Identikit simulation does not
match the structure of the galaxies being modeled?

Suppose for a moment that the Identikit mass model and the real galaxy
have the similar rotation curves but apportion mass differently
between various components.  In this situation the Identikit algorithm
will probably yield good encounter parameters despite the mismatch.
One test is shown in Figs.~\ref{cdf_view} and~\ref{cdf_spin}, where
the results obtained by giving all particles equal weights (red lines)
accurately reproduce the viewing and spin directions of all eight test
systems.  This can be interpreted as an experiment in which real
galaxies with exponential discs (surface density $\Sigma \propto e^{-
\alpha R}$) are matched to Identikit models with radially biased discs
($\Sigma \propto R^2 e^{- \alpha R}$) but identical rotation curves.

At the opposite extreme, suppose an attempt is made to match a pair of
galaxies with long, well-developed tidal tails -- which imply
relatively shallow galactic potential wells -- using an Identikit
model with a very deep potential well.  Since such a model would be
unable to produce long tidal tails, it's unlikely that \textit{any}
set of encounter parameters could populate phase-space regions near
the ends of the tails.  As a result, the algorithm would fail to find
a solution.  This `failure' points to a flaw in the adopted mass
model; the obvious recourse is to try a model with a shallower
potential well.

Between these extremes is a grey area in which Identikit~2 may yield a
good match to the morphology and kinematics of an interacting system
without providing accurate values for all encounter parameters.
Consider the problem of modeling a system observed just after first
encounter, in which tidal features have not yet had time to develop
and probe the full extent of the galactic potential wells.  Some
constraint on potential well depth may still be afforded by the
relative velocities of the two galaxies, but depth is likely to be
degenerate with orbital eccentricity; for example, a large
line-of-sight velocity difference may arise because the galaxies have
deep potential wells \textit{or} because their initial orbit was
hyperbolic ($e > 1$).

What can be learned about galactic structure by matching the
morphology and kinematics of a pair of interacting galaxies with a
dynamical model?  This larger question is independent of the specific
technique -- Identikit~2, genetic algorithm, or trial-and-error --
used to produce the model.  Clearly there's no simple answer; the
outcome will vary from system to system.  Identikit~2 offers a
practical way to explore this question without laborious
trial-and-error modeling.  It's straightforward to construct
self-consistent simulations with various mass models; these could then
be tested against Identikit models with different rotation curves.  Of
particular interest will be tests varying the ratio of circular to
escape velocity, which appears to predict the extent of the tidal
tails produced in an encounter \citetext{\citealp{SW99};
\citealp{DMH99}}.

\subsection{Coda}

Determining additional encounter parameters
(\S~\ref{other_parameters}) and exploring the results of different
mass models (\S~\ref{mass_models}) both present intriguing theoretical
problems.  Beyond these, the effects of `pre-existing conditions' such
as bars and warps suggest additional lines of investigation.  However,
the algorithm already appears quite effective at reconstructing
galactic collisions.  It will be very interesting to see if it works
with real galaxy data.

Source code for this algorithm is available at
\texttt{http://www.ifa.hawaii.edu/faculty/barnes/research/identikit/}.

\section*{Acknowledgments}

I am grateful to John Hibbard for his collaboration in developing
Identikit~1, which led me towards the algorithm presented here.  The
Observatories of the Carnegie Institute of Washington, the California
Institute of Technology, and Kyoto University provided support and
hospitality during the initial stages of this work.  I thank
colleagues at these institutions, and at Academia Sinica, Columbia
University, the Institute for Astronomy, and the National Astronomical
Observatory of Japan for listening to my talks on this project as it
developed and offering useful feedback.  Finally, I thank the referee
for a very positive and constructive report.

\newcommand{\apj}{{ApJ}}
\newcommand{\aj}{{AJ}}
\newcommand{\aap}{{A\&A}}
\newcommand{\aaps}{{A\&AS}}
\newcommand{\mnras}{{MNRAS}}


\begin{thebibliography}{}

\bibitem[Barnes \& Hibbard(2009)]{BH09}
{Barnes, J.E. \& Hibbard, J.E. 2009,
`IDENTIKIT 1: A modeling tool for interacting disc galaxies',
\aj, \textbf{137}, 3071--3090}

\bibitem[Casertano \& van Gorkom(1991)]{CvG91}
{Casertano, S. \& van Gorkom, J.H. 1991, 
`Declining rotation curves -- The end of a conspiracy?', 
\aj, \textbf{101}, 1231--1241}

\bibitem[Catinella, Giovanelli, \& Haynes(2006)]{CGH06}
{Catinella, B., Giovanelli, R., \& Haynes, M.P. 2006, 
`Template Rotation Curves for Disc Galaxies', 
\apj, \textbf{640}, 751--761}

\bibitem[Dubinski, Mihos, \& Hernquist(1996)]{DMH96}
{Dubinski, J., Mihos, J.C., \& Hernquist, L. 1996, 
`Using Tidal Tails to Probe Dark Matter Halos', 
\apj, \textbf{462}, 576--593}

\bibitem[\protect\citeauthoryear{Dubinski, Mihos, \&
Hernquist}{Dubinski et al.}{1999}]{DMH99}
{Dubinski, J., Mihos, J.C., \& Hernquist, L. 1999, 
`Constraining Dark Halo Potentials with Tidal Tails', 
\apj, \textbf{526}, 607--622}

\bibitem[Hernquist(1990)]{H90}
{Hernquist, L.E. 1990,
`An analytical model for spherical galaxies and bulges', 
\apj, \textbf{356}, 359--364}

\bibitem[Hibbard \& Mihos(1995)]{HM95}
{Hibbard, J.E. \& Mihos, J.C. 1995,
`Dynamical Modeling of NGC7252 and the Return of Tidal Material', 
\aj, \textbf{110}, 140--155}

\bibitem[Hibbard et al.(1994)]{HGvGS94}
{Hibbard, J.E., Guhathakurta, P., van Gorkom, J.H., \& Schweizer,
F. 1994,
`Cold, warm, and hot gas in the late-stage merger NGC 7252', 
\aj, \textbf{107}, 67--89}
 
\bibitem[Khochfar \& Burkert(2006)]{KB06}
{Khochfar, S. \& Burkert, A. 2006, 
`Orbital parameters of merging dark matter halos', 
\aap, \textbf{445}, 403--412}

\bibitem[Mo, Mao, \& White(1998)]{MMW98}
{Mo, H.J., Mao, S., \& White, S.D.M. 1998,
`The formation of galactic discs',
\mnras, \textbf{295}, 319--336}

\bibitem[Navarro, Frenk, \& White(1996)]{NFW96}
{Navarro, J.F., Frenk, C.S., \& White, S.D.M. 1996, 
`The Structure of Cold Dark Matter Halos', 
\apj, \textbf{462}, 563--575}

\bibitem[Schweizer(1977)]{S77}
{Schweizer, F. 1977,
`Galaxies with long tails',
in \textit{Structure and Properties of Nearby Galaxies},
eds. E.M. Berkhuijsen \& R. Wielebinski.  D. Reidel, Dordrecht.
p. 279--284}

\bibitem[Springel \& White(1999)]{SW99}
{Springel, V., \& White, S.D.M. 1999, 
`Tidal tails in cold dark matter cosmologies', 
\mnras, \textbf{307}, 162--178}

\bibitem[Stockton(1974)]{S74}
{Stockton, A. 1974, 
`Spectroscopic observations of NGC 4676',
\apj, \textbf{187}, 219--221}

\bibitem[Theis \& Kohle(2001)]{TK01}
{Thies, C. \& Kohle, S. 2001, 
`Multi-method-modeling of interacting galaxies', 
\aap, \textbf{370}, 365--383}

\bibitem[Toomre \& Toomre(1972)]{TT72}
{Toomre, A. \& Toomre, J. 1972, 
`Galactic Bridges and Tails', 
\apj, \textbf{178}, 623--666}

\bibitem[Wahde(1998)]{W98}
{Wahde, M. 1998,
`Determination of orbital parameters of interacting galaxies
using a genetic algorithm. Description of the method and
application to artificial data', 
\aaps, \textbf{132}, 417--429} 
 
\bibitem[Wahde \& Donner(2001)]{WD01}
{Wahde, M. \& Donner, K.J. 2001, 
`Determination of the orbital parameters of the M 51 system using
a genetic algorithm', 
\aap, \textbf{379}, 115--124}

\end{thebibliography}
\end{document}